\documentclass[showpacs,aps,prd,nofootinbib,floatfix,amsmath,amssymb]{revtex4}
\usepackage{graphicx}
\usepackage{dsfont}
\begin{document}

\makeatletter
\newbox\slashbox \setbox\slashbox=\hbox{$/$}
\newbox\Slashbox \setbox\Slashbox=\hbox{\large$/$}
\def\pFMslash#1{\setbox\@tempboxa=\hbox{$#1$}
  \@tempdima=0.5\wd\slashbox \advance\@tempdima 0.5\wd\@tempboxa
  \copy\slashbox \kern-\@tempdima \box\@tempboxa}
\def\pFMSlash#1{\setbox\@tempboxa=\hbox{$#1$}
  \@tempdima=0.5\wd\Slashbox \advance\@tempdima 0.5\wd\@tempboxa
  \copy\Slashbox \kern-\@tempdima \box\@tempboxa}
\def\FMslash{\protect\pFMslash}
\def\FMSlash{\protect\pFMSlash}
\def\miss#1{\ifmmode{/\mkern-11mu #1}\else{${/\mkern-11mu #1}$}\fi}
\makeatother

\title{Distinctive ultraviolet structure of extra-dimensional Yang-Mills theories by integration of heavy Kaluza-Klein modes}

\author{I. Garc\'ia--Jim\'enez$^{(a)}$}
\author{H. Novales-S\' anchez$^{(b)}$}
\author{J. J. Toscano$ {}^{(b,c)}$}
\affiliation{$^{(a)}$Instituto de F\'isica, Benem\'erita Universidad Aut\'onoma de Puebla, Apartado Postal J--48, C. P. 72570 Puebla, Puebla, M\'exico. \\ $^{(b)}$Facultad de Ciencias F\'{\i}sico Matem\'aticas,
Benem\'erita Universidad Aut\'onoma de Puebla, Apartado Postal
1152, Puebla, Puebla, M\'exico.\\
$^{(c)}$Facultad de Ciencias F\'isico Matem\' aticas, Universidad Michoacana de San Nicol\' as de Hidalgo, Avenida Francisco J. M\' ujica S/N,
58060, Morelia, Michoac\' an, M\' exico.}

\begin{abstract}
One--loop Standard Model observables produced by virtual heavy Kaluza--Klein fields play a prominent role in the minimal model of universal extra dimensions. Motivated by this aspect, we integrate out all the Kaluza--Klein heavy modes coming from the Yang--Mills theory set on a spacetime with an arbitrary number, $n$, of compact extra dimensions. After fixing the gauge with respect to the Kaluza--Klein heavy gauge modes in a covariant manner, we calculate a gauge independent effective Lagrangian expansion containing multiple Kaluza--Klein sums that entail a bad divergent behavior.
We use the Epstein--zeta function to regularize and characterize discrete divergences within such multiple sums, and then we discuss the interplay between the number of extra dimensions and the degree of accuracy of effective Lagrangians to generate or not divergent terms of discrete origin. We find that nonrenormalizable terms with mass dimension $k$ are finite as long as $k>4+n$. Multiple Kaluza--Klein sums of nondecoupling logarithmic terms, not treatable by Epstein--zeta regularization, are produced by four--dimensional momentum integration. On the grounds of standard renormalization, we argue that such effects are unobservable.
\end{abstract}

\pacs{11.10.Kk, 11.15.--q, 14.70.Pw,14.80.Rt}

\maketitle
\section{Introduction}
\label{Intro}
The formulation of a physical theory describing nature at the most fundamental level is one of the main incentives behind investigations framed within high--energy physics. While experiments have set some hints~\cite{PDG} pointing towards this aim, on the theoretical side there are many possibilities available. Among the vast collection of ideas, the conjectural existence of compact extra dimensions~\cite{Kclss1,Kclss2,Anto,ADD,AADD} is the setup for the present work.
\\

The exploration of Standard Model extensions with the ingredient of extra dimensions has been motivated by some of the most intriguing questions nowadays. A world in which dark matter particles are part of the field content of Kaluza--Klein effective theories has been widely investigated~\cite{ChMS,ChFM,SerTaotro,SerTa,HooKr,BHS,KoMa,HooPro,DHKM,BNP,BKP,BMMO}. Studies of neutrino physics in extra--dimensional contexts have been carried out as well, including the generation of neutrino masses~\cite{DDGnu,MNPL,MRS,ADDMR,ADPY,BMOZotro,BMOZ,OhRi} and the physics behind neutrino oscillations~\cite{DDGnu,MNPL,ADPY,DvSm,BCS}.
The physics of the Higgs boson in extra--dimensional scenarios has been also an object of study~\cite{AStr,HaKo,RizzWe,FPetr,AppYee,CGM,DLR,BhKu,SaKuRa,BBD,NThiggs,BGHHLMSS,CCGHN,KNOW,BBBKP}.
\\

Among the different models of extra dimensions, there is the formulation in which the whole Standard Model is defined in the extra--dimensional spacetime, where all the dynamic variables are allowed to propagate, so that those particles included by the Standard Model in four dimensions are the lowest--energy manifestations of such extra--dimensional fields, which describe nature at a higher--energy scale. This framework, commonly known as {\it universal extra dimensions}~\cite{ACD}, is the setting of the study performed in the present paper. Investigations centered in the cosmological role of the lightest Kaluza--Klein particle~\cite{BKP,CPS}, a Kaluza--Klein Higgs boson~\cite{BBBKP,DeyRa,DPR}, and experimental data~\cite{ATLASUED,ATLASUED2} from the Large Hadron Collider have provided upper bounds on the size of universal extra dimensions, corresponding to energy scales that range around 1\,TeV for the case of just one compact dimension.
\\

The recent observation~\cite{ATLAShiggs,CMShiggs} of a Higgs--like particle at the Large Hadron Collider is the last piece of the realization that nature is governed by gauge symmetry. In this context, the fate of extra--dimensional gauge symmetry at the level of Kaluza--Klein theories becomes a matter of interest. From the four--dimensional viewpoint, extra--dimensional gauge symmetry is split into two disjoint sets of transformations that have been termed the {\it standard gauge transformations} and the {\it nonstandard gauge transformations}~\cite{NT5DYM,CGNT,LMNT}.
The set of standard gauge transformations is a gauge subgroup that is identified with the gauge symmetry characterizing the four--dimensional low--energy formulation, while the rest of the extra--dimensional gauge group, which corresponds to the nonstandard gauge transformations, remains hidden~\cite{LMNT}.
With these ingredients, the quantization of Kaluza--Klein gauge theories can be carried out~\cite{NT5DYM}.
One can take advantage of the mutual independence of the two four--dimensional sets of gauge transformations and execute quantization while leaving four--dimensional gauge invariance untouched~\cite{NT5DYM}. In practice, the presence of this symmetry is convenient, for it introduces simplifications~\cite{FMNRT,NTKKint} in calculations.
\\

The Becchi--Rouet--Stora--Tyutin quantization~\cite{BRS1,BRS2,Tyutin,GPS} of gauge theories in five dimensions was detailed in a paper by some of us~\cite{NT5DYM}. Among the main points of that work, a four--dimensional set of SU($N,{\cal M}^4$)--covariant gauge--fixing functions was proposed in order to maintain four--dimensional gauge invariance in the quantum Kaluza--Klein theory. This was then utilized to integrate out~\cite{NTKKint} all the Kaluza--Klein excited modes and derive an effective Lagrangian featuring four--dimensional gauge invariance. In the present paper, we go further in this direction and perform such calculation for the case of $n$ extra dimensions. We resort to the results reported in Ref.~\cite{LMNTotro}, in which a full analysis of the Kaluza--Klein Lagrangian generated by the $n$--dimensional Yang--Mills theory was performed.
We find that a gauge independent result can be obtained by the delicate interplay of the contributions from the pure--gauge, pseudo--Goldstone and ghost--antighost Kaluza--Klein sectors.
\\

Nonrenormalizability of extra--dimensional formulations manifests in our results, which include multiple infinite Kaluza--Klein sums.
Using a regularization scheme~\cite{GLMNNT}, which is based on the Epstein--zeta function~\cite{Eps,PoTi,NaPa,Zuck,Hard,Sieg,Glass,Glassotro}, we identify and isolate divergences that are inherent in such multiple sums.
Our effective Lagrangian expansion involves nonrenormalizable terms whose mass--dimension is as large as 6. We determine that Kaluza--Klein sums in these terms produce a divergence if the number of extra dimensions is greater than 1.
Similarly to the general discussion of Ref.~\cite{GLMNNT}, we find that improvements in the accuracy of this effective Lagrangian would modify the ultraviolet behavior, by Kaluza--Klein sums, as long as the number of extra dimensions is large enough. Besides such discrete divergences, there are terms in our effective Lagrangian expansion that include standard divergences (short distance effects on the standard four--dimensional spacetime manifold) and nondecoupling logarithms that are affected by multiple Kaluza--Klein sums. Nevertheless, following previous results~\cite{GLMNNT}, we argue that these nondecoupling effects are unobservable, since they can be  absorbed by the standard renormalization procedure, used to eliminate standard divergences.
\\

A complete and detailed study about the quantization of gauge theories comprising $n$ extra dimensions is underway and will be presented elsewhere~\cite{inprog}. Anyway, in the present paper we provide some advances on that matter, including the tree--level couplings from the ghost--antighost sector that contribute to standard Green's functions\footnote{Throughout the paper we use the term {\it standard Green's function} to refer to any Green's function generated by loop diagrams in which all external legs are Kaluza--Klein zero modes.} at the one--loop level and a generalization to $n$ extra dimensions of the Kaluza--Klein covariant gauge--fixing functions given in Ref.~\cite{NT5DYM}. A simple relation among one--loop contributions from Kaluza--Klein pseudo--Goldstone bosons and those from the ghost--antighost sector is observed, which also occurs in five dimensions~\cite{NTKKint,FMNRT}.
\\

This document has been organized in the following way: we develop a brief discussion on the Kaluza--Klein model in Section~\ref{treeKKto1loop}, which includes the mass--spectrum of  the Kaluza--Klein scalars and the Kaluza--Klein couplings that are necessary to carry out the main calculation; then, in Section \ref{quantasp}, we provide some results on the quantization of the Kaluza--Klein theory; Section~\ref{KKint} is dedicated to the integration of the Kaluza--Klein excited modes, covering the proof of gauge independence and the Epstein--zeta regularization of divergences from Kaluza--Klein sums; finally, in Section~\ref{DandS}, we give a summary of the paper.

\section{Tree--level Kaluza--Klein couplings contributing at one loop}
\label{treeKKto1loop}
Theoretical aspects of field theories in extra dimensions have been considered in diverse works~\cite{ADD,ACD,DDGed1,RS1,RS2,DDGed2,MPR,Hol,DoPo,BDP,Uek,NT5DYM,LMNT,LMNTotro}. In this section, we provide some results that are necessary ingredients to perform the main calculation of the paper. In what follows, we use the notation of Refs.~\cite{LMNTotro,GLMNNT}, where fully detailed discussions on all these results can be found.
\\

We begin by assuming that, at some high--energy scale, spacetime looks like a plane manifold, ${\cal M}^{4+n}$, comprising $4+n$ dimensions and being characterized by a Minkowski--like metric, $g^{MN}={\rm diag}(1,-1,\ldots,-1)$. Uppercase indices run over all the spacetime coordinates, so that $M,N=0,1,2,3,5,\ldots,4+n$.
Any field formulation nested in this spacetime will be governed by the extra--dimensional Poincar\'e group ISO($1,3+n$). We also assume that all fields propagate in the whole spacetime, so that they are functions of $(4+n)$--vector coordinates $(x,\bar{x})=(x^0,x^1,x^2,x^3,\bar{x}^5,\ldots,\bar{x}^{4+n})$.
We consider the SU($N,{\cal M}^{4+n}$)--invariant Lagrangian
\begin{equation}
{\cal L}_{\rm YM}(x,\bar{x})=-\frac{1}{4}{\cal F}^a_{MN}(x,\bar{x})\,{\cal F}^{aMN}(x,\bar{x}),
\label{5dYML}
\end{equation}
given in terms of extra--dimensional gauge fields, which are denoted by ${\cal A}^a_M(x,\bar{x})$ and which define the Yang--Mills curvatures as ${\cal F}^a_{MN}(x,\bar{x})=\partial_M{\cal A}^a_N(x,\bar{x})-\partial_N{\cal A}^a_M(x,\bar{x})+g_{4+n}\,f^{abc}{\cal A}^b_M(x,\bar{x})\,{\cal A}^c_N(x,\bar{x})$. Here, the $g_{4+n}$ is the SU($N,{\cal M}^{4+n}$) coupling constant, with dimensions $({\rm mass})^{-n/2}$, and $f^{abc}$ represents the structure constants of the gauge group. Lowercase indices correspond to the gauge group, which means that $a=1,2,\ldots,N^2-1$.
\\

So far, experiments have not found any evidence~\cite{PDG,ATLASUED,ATLASUED2} pointing to the actual existence of extra dimensions, which can be explained as long as these extra dimensions are small enough.
The transition of the ${\cal L}_{\rm YM}(x,\bar{x})$ Lagrangian from the extra--dimensional manifold ${\cal M}^{4+n}$ to the four--dimensional perspective is implemented~\cite{GLMNNT} by two canonical transformations.
The first of such transformations maps covariant objects of ISO($1,3+n$) into ISO($1,3$)--covariant objects:
extra--dimensional vector gauge fields ${\cal A}^a_M(x,\bar{x})$ are split into ISO($1,3$) 4--vectors ${\cal A}^a_\mu(x,\bar{x})$, with $\mu=0,1,2,3$, and a set of ISO($1,3$) scalars ${\cal A}^a_{\bar{\mu}}(x,\bar{x})$, in which $\bar{\mu}=5,6,\ldots,4+n$.
This transformation maintains invariance under the extra--dimensional Poincar\'e group, though keeping it hidden and only showing manifest invariance with respect to ISO(1,3). As we consider lower energies, the compact structure of extra dimensions becomes apparent and the extra--dimensional Poincar\'e group ISO$(1,3+n)$ is explicitly broken by compactification. We assume that the resulting spacetime is ${\cal M}^4\times{\cal N}^n$, where ${\cal M}^4$ represents the standard four--dimensional spacetime and the submanifold ${\cal N}^n=(S^1/Z_2)^n$ stands for all compact dimensions, whose radii are $R_1, R_2,\ldots, R_n$. With this structure of the compact extra dimensions, the gauge fields ${\cal A}^a_M(x,\bar{x})$ acquire periodicity properties, ${\cal A}^a_M(x,\bar{x})={\cal A}^a_M(x,\bar{x}+2\pi R)$, and parity properties, ${\cal A}^a_M(x,-\bar{x})=\pm{\cal A}^a_{M}(x,\bar{x})$, as well.
The implementation in the ${\cal L}_{\rm YM}(x,\bar{x})$ Lagrangian of the explicit breaking of the extra--dimensional Poincar\'e group ISO(1,$3+n$) is carried out by~\cite{LMNT,GLMNNT} a second canonical transformation, which turns out to be Fourier expansions of the ${\cal A}^a_M(x,\bar{x})$ fields that are consistent with the periodicity and parity properties adopted by them. These expansions are commonly known as {\it Kaluza--Klein towers}.
\\

After using the aforementioned Fourier series, all dependence on the extra--dimensional coordinates of the Yang--Mills Lagrangian ${\cal L}_{\rm YM}(x,\bar{x})$, Eq.~(\ref{5dYML}), can be integrated out, which generates a four--dimensional effective theory, ${\cal L}_{\rm 4YM}(x)$, whose dynamic variables are the Kaluza--Klein modes. To make sure that the low--energy limit of ${\cal L}_{\rm 4YM}(x)$ is just the Yang--Mills theory in four dimensions the correct parity conditions for the extra--dimensional fields are ${\cal A}^a_\mu(x,-\bar{x})=+{\cal A}^a_\mu(x,\bar{x})$, and ${\cal A}^a_{\bar{\mu}}(x,-\bar{x})=-{\cal A}^a_{\bar{\mu}}(x,\bar{x})$. With this choice, the $A^a_{\mu}(x,\bar{x})$ 4--vector unfolds into a set of $2^n(N^2-1)$ gauge Kaluza--Klein modes, of which ($N^2-1$) are zero modes, $A^{(0,\ldots,0)a}_\mu(x)$, and $(2^n-1)(N^2-1)$ are excited modes, $A^{(k_1,\ldots,k_n)a}_\mu(x)$.
For excited modes, the {\it Kaluza--Klein indices} $k_1,\ldots,k_n$ are nonnegative integer numbers, but the case in which all of these indices are simultaneously zero is excluded, for it corresponds to zero modes.
The zero modes are recognized as the dynamic variables of the low--energy theory, and consistently they behave as gauge fields with respect to the standard gauge transformations.
The ISO(1,3) scalars ${\cal A}^a_{\bar{\mu}}(x,\bar{x})$, on the other hand, are decomposed into $n\,(2^n-1)(N^2-1)$ scalar Kaluza--Klein modes, $A^{(k_1,\ldots,k_n)a}_{\bar{\mu}}(x)$. Again, the only combination of Kaluza--Klein indices that is excluded is $(0,\ldots,0)$.
\\

The general case of $n$ extra dimensions brings intricate expressions, which are difficult to read. For this reason, a convenient notation, which evokes intuition and directly generalices the results of the five--dimensional case~\cite{NT5DYM} is desirable. To this aim, we define $(\underline{0})=(0,\ldots,0)$, which we use to express zero modes more briefly as $A_\mu^{(0,\ldots,0)a}(x)=A^{(\underline{0})a}_\mu(x)$. We represent all other possible arrangements of Kaluza--Klein indices $(k_1,0,\ldots,0),\ldots,(0,\ldots,k_n),$ $(k_1,k_2,0\ldots,0),\ldots,\,(0,\ldots,0,k_{n-1},k_n),\ldots,$\,$(k_1,\ldots,k_n)$ generically by $(\underline{k})$, so that Kaluza--Klein excited modes are compactly denoted by $A^{(\underline{k})a}_\mu(x)$ and $A^{(\underline{k})a}_{\bar{\mu}}(x)$. Then, we use the sum~\cite{LMNTotro,GLMNNT}
\begin{eqnarray}
\sum_{(\underline{k})}f^{(\underline{k})}&=&\sum_{k_1=1}^\infty f^{(k_1,0,\ldots,0)}+\cdots+\sum_{k_n=1}^\infty f^{(0,\ldots,0,k_n)}
+\sum_{k_1=1}^\infty\sum_{k_2=1}^\infty f^{(k_1,k_2,0,\ldots,0)}
\nonumber \\ \nonumber \\ &&
+\cdots+\sum_{k_{n-1}=1}^\infty\sum_{k_n=1}^\infty f^{(0,\ldots,0,k_{n-1},k_n)}+\cdots+\sum_{k_1=1}^\infty\cdots\sum_{k_n=1}^\infty f^{(k_1,\ldots,k_n)},
\label{KKsumdef}
\end{eqnarray}
which comprises all possible arrangements of Kaluza--Klein indices for a given number of extra dimensions. In terms of the sum given in Eq.~(\ref{KKsumdef}), all results have the same structure that is found in the case of just one extra dimension~\cite{NT5DYM}. Notice that this definition involves multiple infinite sums.
\\

The extra--dimensional curvature ${\cal F}^a_{MN}(x,\bar{x})$ inherits, from the gauge fields ${\cal A}^a_M(x,\bar{x})$ defining it, specific periodicity and parity transformation properties with respect to the extra coordinates, which means that it can also be expanded in Kaluza--Klein towers. The implementation of the first canonical transformation in the extra--dimensional Lagrangian ${\cal L}_{\rm YM}(x,\bar{x})$ separates the extra--dimensional curvature into three components that possess definite transformation properties under ISO($1,3$), that is, ${\cal F}_{MN}(x,\bar{x})\to{\cal F}_{\mu\nu}(x,\bar{x}),\,{\cal F}_{\mu\bar{\nu}}(x,\bar{x}),\,{\cal F}_{\bar{\mu}\bar{\nu}}(x,\bar{x})$, with ${\cal F}_{\mu\nu}(x,\bar{x})$ being a 2--tensor, ${\cal F}_{\mu\bar{\nu}}(x,\bar{x})$ transforming as a vector, and ${\cal F}_{\bar{\mu}\bar{\nu}}(x,\bar{x})$ behaving as an ISO(1,3) scalar. The second canonical transformation, corresponding to the Kaluza--Klein towers, produces a set of Kaluza--Klein excitations for each of these components, which are given in terms of the Kaluza--Klein modes of the extra--dimensional gauge fields as~\cite{LMNTotro}
\begin{eqnarray}
{\cal F}^{(\underline{0})a}_{\mu\nu}&=&F^{(\underline{0})a}_{\mu\nu}+gf^{abc}\sum_{(\underline{k})}A^{(\underline{k})b}_\mu A^{(\underline{k})c}_\nu,
\label{zmgc}
\\ \nonumber \\
{\cal F}^{(\underline{0})a}_{\bar{\mu}\bar{\nu}}&=&gf^{abc}\sum_{(\underline{k})}A^{(\underline{k})b}_{\bar{\mu}}A^{(\underline{k})c}_{\bar{\nu}},
\\ \nonumber \\
{\cal F}^{(\underline{n})a}_{\mu\nu}&=&{\cal D}^{(\underline{0})ab}_{\mu}A^{(\underline{n})b}_{\nu}-{\cal D}^{(\underline{0})ab}_{\nu}A^{(\underline{n})b}_{\mu}+gf^{abc}\sum_{(\underline{kr})}\Delta_{(\underline{nkr})}\,A^{(\underline{k})b}_\mu A^{(\underline{r})c}_\nu,
\\ \nonumber \\
{\cal F}^{(\underline{n})a}_{\mu\bar{\nu}}&=&{\cal D}_\mu^{(\underline{0})ab} A^{(\underline{n})b}_{\bar{\nu}}+p^{(\underline{n})}_{\bar{\nu}}\,A^{(\underline{n})a}_\mu+gf^{abc}\sum_{(\underline{kr})}\Delta'_{(\underline{nrk})}\,A^{(\underline{k})b}_\mu A^{(\underline{r})c}_{\bar{\nu}},
\\ \nonumber \\
{\cal F}^{(\underline{n})a}_{\bar{\mu}\bar{\nu}}&=&p^{(\underline{n})}_{\bar{\mu}}A^{(\underline{n})a}_{\bar{\nu}}-p^{(\underline{n})}_{\bar{\nu}}A^{(\underline{n})a}_{\bar{\mu}}+gf^{abc}\sum_{(\underline{kr})}\Delta'_{(\underline{krn})}\,A^{(\underline{k})b}_{\bar{\mu}}A^{(\underline{r})c}_{\bar{\nu}}.
\end{eqnarray}
where $F^{(\underline{0})a}_{\mu\nu}=\partial_\mu A^{(\underline{0})a}_\nu-\partial_\nu A^{(\underline{0})a}_\mu+g f^{abc}A^{(\underline{0})b}_\mu A^{(\underline{0})c}_\nu$ is the four--dimensional Yang--Mills curvature and ${\cal D}^{(\underline{0})ab}_\mu=\delta^{ab}\partial_\mu-gf^{abc}A^{(\underline{0})c}_\mu$ is the SU$(N,{\cal M}^4)$ covariant derivative. Both of these objects include the SU$(N,{\cal M}^4)$ coupling constant, $g$, which is related to its extra--dimensional counterpart by $g=g_{4+n}/\sqrt{(2\pi R_1)\cdots(2\pi R_n) }$. In addition, we have defined
\begin{equation}
p^{(\underline{k})}_{\bar{\mu}}=\sum_{\alpha=1}^n\frac{\underline{k}_\alpha}{R_\alpha}\delta_{\bar{\mu}\,4+\alpha},
\end{equation}
with the underlining of $\underline{k}_\alpha$ indicating that $k_\alpha$ can be either zero or a natural number, which depends on the zero and nonzero Kaluza--Klein indices in any concrete combination $(\underline{k})$ that we take in $m_{\bar{\mu}}(\underline{k})$.
Some of these Kaluza--Klein modes include the objects $\Delta_{(\underline{nrk})}$ and $\Delta'_{(\underline{nrk})}$. While their precise definitions, given in terms of Kronecker deltas, can be found in Ref.~\cite{LMNTotro}, it is worth commenting that in what follows we do not consider any term in which they appear. The reason is that couplings that incorporate these objects do not contribute at the one--loop level to standard Green's functions, but they do it since higher orders.
\\

It turns out that the Kaluza--Klein Lagrangian is expressed in terms of the Kaluza--Klein modes of the curvature in a relatively simple manner. The precise expression reads~\cite{LMNTotro}
\begin{equation}
{\cal L}_{\rm 4YM}=-\frac{1}{4}\Bigg[ {\cal F}^{(\underline{0})a}_{\mu\nu}{\cal F}^{(\underline{0})a\mu\nu}+{\cal F}^{(\underline{0})a}_{\bar{\mu}\bar{\nu}}{\cal F}^{(\underline{0})a\bar{\mu}\bar{\nu}}+\sum_{(\underline{k})}\Big( {\cal F}^{(\underline{k})a}_{\mu\nu}{\cal F}^{(\underline{k})a\mu\nu}
+2 {\cal F}^{(\underline{k})a}_{\mu\bar{\nu}}{\cal F}^{(\underline{k})a\mu\bar{\nu}}+{\cal F}^{(\underline{k})a}_{\bar{\mu}\bar{\nu}}{\cal F}^{(\underline{k})a\bar{\mu}\bar{\nu}} \Big) \Bigg].
\label{KKLcurv}
\end{equation}
In the next two subsections we extract from Eq.~(\ref{KKLcurv}) all those Kaluza--Klein couplings that contribute to standard Green's functions at one loop.

\subsection{Kaluza--Klein scalars and mass spectrum}
The emergence of Kaluza--Klein scalar modes, after compactification, is an interesting characteristic of the SU($N,{\cal M}^{4+n}$) theory. In the case of just one extra dimension, the number of Kaluza--Klein scalars $A^{(k)a}_5(x)$ is $N^2-1$, which exactly matches the number of Kaluza--Klein gauge modes $A^{(k)a}_\mu(x)$. Remarkably, the gauge excited modes are massive, even though they originate from massless five--dimensional gauge fields. The scalar Kaluza--Klein modes behave  like pseudo--Goldstone bosons in the sense that they are massless and can be eliminated from the theory by a specific gauge transformation~\cite{NT5DYM}, just like if they had given their physical degrees of freedom to the excited Kaluza--Klein gauge fields. With the assumption of more extra dimensions the analysis grows in difficulty, although the same mechanism for generation of gauge masses\footnote{By {\it gauge masses} we mean that the corresponding mass terms are invariant under the standard gauge group $SU(N,{\cal M}^4)$.} takes place~\cite{LMNTotro,GLMNNT}. Some complications in the scalar sector arise because for $n>1$ extra dimensions the number of Kaluza--Klein scalars is greater than the number of Kaluza--Klein gauge modes.  Another difficulty is introduced by the presence of mixings among some of the scalar Kaluza--Klein modes. The whole set of $n(2^n-1)(N^2-1)$ Kaluza--Klein scalars can be split into two types of fields, according to whether or not they participate in such scalar mixings: $n\,2^{n-1}(N^2-1)$ Kaluza--Klein scalars mix, whereas $n(2^{n-1}-1)(N^2-1)$ do not take part in mixings.
All this information is enclosed by the fifth term of the right--hand side of Eq.~(\ref{KKLcurv}).
\\

The fifth term of Eq.~(\ref{KKLcurv}) can be written as
\begin{eqnarray}
-\frac{1}{4}\sum_{(\underline{k})}{\cal F}^{(\underline{k})a}_{\bar{\mu}\bar{\nu}}{\cal F}^{(\underline{k})a}_{\bar{\mu}\bar{\nu}}&=&-\frac{1}{2}\sum_{(\underline{k})}A^{(\underline{k})a}_{\bar{\mu}}\,\mathfrak{M}_{\bar{\mu}\bar{\nu}}^{(\underline{k})}\,A^{(\underline{k})a}_{\bar{\nu}}+\cdots.
\label{Lsmass}
\end{eqnarray}
The only term explicitly shown in the right--hand side of the last equation comprises a set of $n\times n$ mixing matrices, $\mathfrak{M}^{(\underline{k})}_{\bar{\mu}\bar{\nu}}$, each of which corresponds to a fixed combination of Kaluza--Klein indices in the sum over $(\underline{k})$. All the information concerning Kaluza--Klein scalar mixings is contained in these matrices, with components concisely expressed as
\begin{equation}
\mathfrak{M}^{(\underline{k})}_{\bar{\mu}\bar{\nu}}=m^2_{(\underline{k})}\delta_{\bar{\mu}\bar{\nu}}
-p^{(\underline{k})}_{\bar{\mu}}\,p^{(\underline{k})}_{\bar{\nu}}.
\label{Mmixingcomps}
\end{equation}
This is the structure of the inertia tensor associated to a single massive particle located at ${\bf r}^{\rm T}(\underline{k})=(p^{(\underline{k})}_5,\ldots,p^{(\underline{k})}_{4+n})$. The shape of any mixing matrix is determined by the number of nonzero Kaluza--Klein indices in the combination $(\underline{k})$ that distinguishes it: the number of mixed Kaluza--Klein scalars matches the number of nonzero Kaluza--Klein indices, whereas all the remaining, and unmixed, scalars have definite mass,
\begin{equation}
m_{(\underline{k})}=\sqrt{\left( \frac{\underline{k}_1}{R_1} \right)^2+\cdots+\left( \frac{\underline{k}_n}{R_n} \right)^2},
\end{equation}
from the onset. Indeed, by performing appropriate interchanges of columns and rows in any mixing matrix $\mathfrak{M}^{(\underline{k})}_{\bar{\mu}\bar{\nu}}$, with $r$ nonzero Kaluza--Klein indices, it can be rearranged as a block matrix that looks like
\begin{equation}
\mathfrak{M}^{(\underline{k})}_{\bar{\mu}\bar{\nu}}\longrightarrow
\left(
\begin{array}{cc}
m_{(\underline{k})}^2\cdot{\bf 1}_{n-r}^{(\underline{k})} & 0
\\ \\
0 & {\cal M}^{(\underline{k})}_r
\end{array}
\right),
\label{blockmixingmatrix}
\end{equation}
where ${\bf 1}^{(\underline{k})}_{n-r}$, in the block $m^2_{(\underline{k})}\cdot{\bf 1}^{(\underline{k})}_{n-r}$, is the $(n-r)\times(n-r)$ identity matrix and ${\cal M}^{(\underline{k})}_r$ is an $r\times r$ nondiagonal matrix that mixes $r$ Kaluza--Klein scalars. It is worth emphasizing that, while any scalar mixing matrix can be manipulated to fit this generic shape, the sizes of the ${\bf 1}^{(\underline{k})}_{n-r}$ and ${\cal M}^{(\underline{k})}_r$ matrices depend on the number of nonzero Kaluza--Klein indices, so that the precise structure of each $\mathfrak{M}^{(\underline{k})}$ clearly depends on its corresponding combination $(\underline{k})$. 
A detailed description of the specific mixing pattern followed by the set of Kaluza--Klein scalars has been carried out in Ref.~\cite{LMNTotro}.
\\

Of course, scalar mixing can be eradicated from the Kaluza--Klein theory by diagonalizing the mixing matrices $\mathfrak{M}^{(\underline{k})}_{\bar{\mu}\bar{\nu}}$. All such matrices have the same eigenvalue spectrum: 1 zero eigenvalue and $n-1$ nonzero eigenvalues, all of them being equal to $m^2_{(\underline{k})}$. This corresponds to a mass--eigenstates basis characterized by 1 massless scalar, $A^{(\underline{k})a}_G$, and $n-1$ massive scalars, $A^{(\underline{k})a}_{\bar{n}}$, with $\bar{n}=1,2,\ldots,n-1$. For $n$ extra dimensions, the total number of scalar mixings in the Kaluza--Klein theory is $(2^n-1)(N^2-1)$, each one providing one massless scalar $A^{(\underline{k})a}_G$. Hence, the total number of massless scalars is $(2^n-1)(N^2-1)$, which consistently coincides with the number of gauge Kaluza--Klein excited modes $A^{(\underline{k})a}_\mu$. To each mixing matrix $\mathfrak{M}^{(\underline{k})}_{\bar{\mu}\bar{\nu}}$ there corresponds a rotation matrix ${\cal R}^{(\underline{k})}_{\bar{\mu}\bar{\mu}'}$ such that
\begin{equation}
\mathfrak{M}^{(\underline{k})}_{\bar{\mu}\bar{\nu}}={\cal R}^{(\underline{k})}_{\bar{\mu}\bar{\mu}'}\,\mathfrak{M}^{(\underline{k})}_{\bar{\mu}'\bar{\nu}'}\,{\cal R}^{(\underline{k})}_{\bar{\nu}'\bar{\nu}}.
\end{equation}
In our notation, any matrix $\mathfrak{M}^{(\underline{k})}$ with primed indices $\bar{\mu}',\bar{\nu}'\,(=1,2,\ldots,n-1,G)$ is a diagonal matrix, contrastingly to matrices with unprimed indices $\bar{\mu},\bar{\nu}\,\,(=5,6,\ldots,4+n)$, which are nondiagonal. Things can always be accommodated in such a manner that, after diagonalization, the last entry of any resulting diagonal matrix $\mathfrak{M}^{(\underline{k})}_{\bar{\mu}'\bar{\nu}'}$ is the zero eigenvalue, which means that $\mathfrak{M}^{(\underline{k})}_{\bar{\mu}'\bar{\nu}'}={\rm diag}(m^2_{(\underline{k})},\ldots,m^2_{(\underline{k})},0)$. This allows a straightforward extraction of the massless scalars $A^{(\underline{k})a}_G$ from
\begin{equation}
A^{(\underline{k})a}_{\bar{\mu}}={\cal R}^{(\underline{k})}_{\bar{\mu}\bar{\mu}'}\,A^{(\underline{k})a}_{\bar{\mu}'}={\cal R}^{(\underline{k})}_{\bar{\mu}\bar{n}}\,A^{(\underline{k})a}_{\bar{n}}+{\cal R}^{(\underline{k})}_{\bar{\mu}G}\,A^{(\underline{k})a}_G,
\end{equation}
to finally express the scalar--mass terms in Eq.~(\ref{Lsmass}) as
\begin{equation}
-\frac{1}{4}\sum_{(\underline{k})}{\cal F}^{(\underline{k})a}_{\bar{\mu}\bar{\nu}}{\cal F}^{(\underline{k})a}_{\bar{\mu}\bar{\nu}}=-\frac{1}{2}\sum_{(\underline{k})}\,m^2_{(\underline{k})}\,A^{(\underline{k})a}_{\bar{n}}A^{(\underline{k})a}_{\bar{n}}-\frac{1}{2}\sum_{(\underline{k})}\,0\times A^{(\underline{k})a}_G\,A^{(\underline{k})a}_G+\cdots.
\end{equation}

\subsection{Kaluza--Klein couplings}
Now that we have discussed the pure--scalar sector of the Kaluza--Klein theory, we proceed to set apart from ${\cal L}_{\rm 4YM}$ all those couplings that contribute to standard Green's functions through one--loop diagrams and which are thus necessary for the integration of the Kaluza--Klein excited modes. An exhaustive catalog of Kaluza--Klein couplings, in the general context of the full extra--dimensional Standard Model, can be found in Ref~\cite{GLMNNT}.
\\

From the expression of the zero mode ${\cal F}^{(\underline{0})a}_{\mu\nu}$, exhibited in Eq.~(\ref{zmgc}), it is clear that the first term of Eq.~(\ref{KKLcurv}) includes the four--dimensional Yang--Mills Lagrangian, defined in terms of the four--dimensional Yang--Mills curvature $F^{(\underline{0})a}_{\mu\nu}$ and which we denote by ${\cal L}^{(\underline{0})}_{\rm 4YM}$. We express this term as
\begin{equation}
-\frac{1}{4}{\cal F}^{(\underline{0})a}_{\mu\nu}{\cal F}^{(\underline{0})a\mu\nu}={\cal L}^{(\underline{0})}_{\rm 4YM}+\frac{1}{2}\,gf^{abc}\sum_{(\underline{k})}A^{(\underline{k})a\mu}F^{(\underline{0})b}_{\mu\nu}A^{(\underline{k})c\nu}+\cdots.
\end{equation}
Besides the low--energy theory, this equation shows explicitly a term contributing at one loop to standard Green's functions. Ellipsis, on the other hand, represents couplings whose lowest--order contributions to such Green's functions enter at the two--loop level. Another term generating pure--gauge interactions within ${\cal L}_{\rm 4YM}$ is the third one, which we write as
\begin{eqnarray}
-\frac{1}{4}\sum_{(\underline{k})}{\cal F}^{(\underline{k})a}_{\mu\nu}{\cal F}^{(\underline{k})a\mu\nu}&=&\sum_{(\underline{k})}\bigg[\,\frac{1}{2}\,g_{\mu\nu}A^{(\underline{k})a\mu}{\cal D}^{(\underline{0})ab}_\rho{\cal D}^{(\underline{0})bc\rho} A^{(\underline{k})c\nu}-\frac{1}{2}\,A^{(\underline{k})a\mu}{\cal D}^{(\underline{0})ab}_\mu{\cal D}^{(\underline{0})bc}_\nu A^{(\underline{k})c\nu}
\nonumber \\ &&
+\frac{1}{2}\,gf^{abc}A^{(\underline{k})a\mu}F^{(\underline{0})b}_{\mu\nu}A^{(\underline{k})c\nu}\bigg]+\cdots.
\end{eqnarray}
\\

The second term of Eq.~(\ref{KKLcurv}) produces only quartic interactions of Kaluza--Klein excited modes, so we omit it and pass to the Kaluza--Klein gauge--scalar interactions that are situated in the fourth term of this equation. This term can be expressed as
\begin{eqnarray}
\frac{1}{2}\sum_{(\underline{k})}{\cal F}^{(\underline{k})a}_{\mu\bar{\nu}}{\cal F}^{(\underline{k})a\mu}\hspace{0.00001cm}_{\bar{\nu}}&=&\sum_{(\underline{k})}\bigg[-\frac{1}{2}A^{(\underline{k})a}_{\bar{\nu}}{\cal D}^{(\underline{0})ab}_\rho{\cal D}^{(\underline{0})bc\rho}A^{(\underline{k})c}_{\bar{\nu}}+p^{(\underline{k})}_{\bar{\nu}}\,A^{(\underline{k})a}_\mu{\cal D}^{(\underline{0})ab\mu}A^{(\underline{k})b}_{\bar{\nu}}
\nonumber \\ \nonumber \\  &&
+\frac{1}{2}\,m_{(\underline{k})}^2\,g_{\mu\nu}A^{(\underline{k})a\mu}A^{(\underline{k})a\nu}\,\bigg]+\cdots.
\label{gsintcurv}
\end{eqnarray}
As the third term of this equation shows, the Kaluza--Klein gauge--scalar sector includes $(2^n-1)(N^2-1)$ mass terms for the whole set of Kaluza--Klein gauge excited modes $A^{(\underline{k})a}_\mu$.
The scalar fields $A^{(\underline{k})a}_{\bar{\nu}}$ in the first term of Eq.~(\ref{gsintcurv}) can be directly rotated into the mass--eigenstate fields $A^{(\underline{k})a}_{\bar{n}}$ and $A^{(\underline{k})a}_G$, just by using orthogonality of ${\cal R}^{(\underline{k})}_{\bar{\mu}\bar{\mu}'}$.
An interesting feature of Eq.~(\ref{gsintcurv}) is the presence, in its second term, of Kaluza--Klein gauge--scalar couplings involving the SU($N,{\cal M}^4$) covariant derivative.
It turns out that the relation
\begin{equation}
p^{(\underline{k})}_{\bar{\nu}}{\cal R}^{(\underline{k})}_{\bar{\nu}\bar{\nu}'}=m_{(\underline{k})}\delta_{\bar{\nu}'G}
\end{equation}
holds for any combination $(\underline{k})$. This yields an exact cancellation of most gauge--scalar couplings in this term, in which the only Kaluza--Klein scalars that survive the rotation ${\cal R}^{(\underline{k})}_{\bar{\nu}\bar{\nu}'}$, and thus participate in such gauge--scalar couplings, are the pseudo--Goldstone bosons $A_G^{(\underline{k})a}$.
Before compactification, the interactions among components of the extra--dimensional gauge vector field ${\cal A}^a_M(x,\bar{x})$ are explicitly governed by the extra--dimensional SU($N,{\cal M}^{4+n}$) gauge symmetry group.
Once the compactness of extra dimensions is implemented in the Lagrangian by the aforementioned canonical transformations, and extra--dimensional gauge invariance is hidden, there emerge the Kaluza--Klein excited gauge modes $A^{(\underline{k})a}_\mu$ and the scalar modes $A^{(\underline{k})a}_{\bar{\mu}}$ as well. After the explicit breaking of the ISO($1,3+n$) group takes place, the couplings of extra--dimensional gauge fields evolve into the couplings and mass terms characterizing the four--dimensional formulation. In particular, a link between gauge excited modes $A^{(\underline{k})a}_\mu$ and a subset of the Kaluza--Klein scalar spectrum is developed. Such link, which manifests through the generation of gauge masses for the gauge excited modes $A^{(\underline{k})a}_\mu$ and the emergence of nonphysical scalars $A^{(\underline{k})a}_G$, also selectively allows the presence of bilinear gauge--scalar couplings: the only Kaluza--Klein scalars that bilinearly couple to Kaluza--Klein gauge modes are the pseudo--Goldstone bosons $A^{(\underline{k})a}_G$, while such interactions are exactly eliminated for the rest of the scalar spectrum.
As we show later, a convenient set of gauge--fixing functions allows us to trade the only existing gauge--scalar bilinear couplings by gauge--dependent mass terms for the pseudo--Goldstone bosons $A^{(\underline{k})a}_G$.

\section{Aspects of quantization}
\label{quantasp}
Gauge symmetry is a profound concept~\cite{Dir,Sunder,GiTy,HeTe} that characterizes successful and accurate physical formulations, realized within field theory, that are aimed to the quantum description of nature. The essence of gauge invariance is the incorporation of more degrees of freedom than those which are strictly necessary to describe a given physical system. While this symmetry manifests as the invariance of Lagrangians under gauge transformations, the Dirac's algorithm~\cite{Dir} dives into the depths of this concept and even provides tools to determine~\cite{Castel} the corresponding gauge transformations. This instrument was used in Refs.~\cite{NT5DYM,LMNT} to develop a careful and complete study of gauge symmetry in the context of extra--dimensional gauge theories. Kaluza--Klein effective descriptions that originate in gauge extra--dimensional theories are invariant under two disjoint sets of gauge transformations: the standard gauge transformations, with respect to which the zero modes $A^{(\underline{0})a}_\mu$ behave as gauge fields; and the nonstandard gauge transformations, which transform the Kaluza--Klein excited vector modes $A^{(\underline{k})a}_\mu$ as gauge fields.
While there are two types of transformations that are independent of each other, it is indeed the full extra--dimensional gauge group the one which governs the interactions of the Kaluza--Klein Lagrangian. Nevertheless, the sets of canonical transformations that take the extra--dimensional Lagrangian ${\cal L}_{\rm YM}$ into the Kaluza--Klein theory ${\cal L}_{\rm 4YM}$ hide~\cite{LMNT} gauge symmetry living in extra dimensions,
in such a way that the ordinary four--dimensional world displays explicit invariance only under the SU($N,{\cal M}^4$) group.
\\

Though gauge symmetry is usually evoked to construct models, the quantization process by path integral requires~\cite{HeTe} this overdescription to be removed. The framework to execute the quantization of gauge systems is provided by the field--antifield formalism~\cite{GPS,FA1,FA2,FA3,FA4,FA5} and the Becchi--Rouet--Stora--Tyutin symmetry~\cite{BRS1,BRS2,Tyutin}, better known as the BRST symmetry. The main point is the determination of a proper solution to the master equation, which arises after a series of extensions of the field spectrum are carried out. In particular, the ghost and antighost fields are introduced, and a set of auxiliary fields enter the game as well. The resulting set of fields is then doubled by introducing an antifield per each field, and a symplectic structure, known as the antibracket, is defined. The proper solution turns out to be the generator of the BRST transformations, which include, as a particular case, the gauge transformations.
In this context, the fixation of the gauge, intended to remove all degeneracy associated to gauge symmetry, is carried out in a nontrivial manner.
This is accomplished by defining a fermionic functional that is used to eliminate all the aforementioned antifields and collaterally fix the gauge.
The main outcome of this procedure is the derivation of a quantum action that depends on general gauge--fixing functions. At this level, gauge symmetry is no longer present and the system is properly quantized.
\\

The quantization of Yang--Mills theories in five spacetime dimensions has been carried out~\cite{NT5DYM} in this approach. The simplest strategy~\cite{NT5DYM,inprog} consists in generalizing the well--known proper solution that corresponds~\cite{GPS} to the four--dimensional version of this formulation to the case in which there exist extra dimensions.
The transition to the four--dimensional perspective gives rise to a richer Kaluza--Klein theory that now includes, besides the ${\cal L}_{\rm 4YM}$ Lagrangian, gauge--fixing and ghost--antighost sectors.
Since the two coexisting sets of four--dimensional gauge transformations are independent of each other, it is possible to remove only~\cite{NT5DYM,inprog} invariance under the nonstandard gauge transformations. In such manner, the four--dimensional SU($N,{\cal M}^4$) symmetry is still valid and the zero modes $A^{(\underline{0})a}_\mu$ are still gauge fields, similarly to what happens, for instance, with the background field method~\cite{DeWitt,DeWittbook,IPS,tH,GNW,KluZu,Boul,Hart,Abbott1,Abbott2,AGS}. In practice, this is achieved by introducing~\cite{NT5DYM,inprog} an {\it ad hoc} set of gauge--fixing functions that are SU($N,{\cal M}^4$)--covariant. This {\it modus operandi} to fix the gauge has been of benefit in phenomenological calculations~\cite{MTTR} framed within other formulations, such as the 331 model~\cite{331PP,331F}.
In a forthcoming paper~\cite{inprog}, this picture will be discussed in full detail. However, since the present paper requires some results concerning  the quantum version of the Kaluza--Klein theory from Yang--Mills in $n$ extra dimensions, we provide here all indispensable expressions for the main calculation.
\\

In the $(4+n)$--dimensional case, the quantization procedure sketched above generates a quantum Kaluza--Klein Lagrangian, ${\cal L}_{\rm QKK}$, that can be split into a sum of three parts as ${\cal L}_{\rm QKK}={\cal L}_{\rm 4YM}+{\cal L}_{\rm GF}+{\cal L}_{\rm G}$, where ${\cal L}_{\rm GF}$ is the gauge--fixing term, defined completely by the gauge--fixing functions, here denoted by $f^{(\underline{k})a}$. The ${\cal L}_{\rm G}$ term represents the sector of ghost and antighost fields, and is determined in part by the election of the gauge--fixing functions. Different sets of these functions have been propounded~\cite{DDGed2,MPR,Uek} for the case of just one extra dimension. We generalize the proposal of Ref.~\cite{NT5DYM}, given for the framework of one extra dimension, and provide the following set of SU($N,{\cal M}^4$)--covariant gauge--fixing functions, which is suitable for $n$ extra dimensions:
\begin{eqnarray}
f^{(\underline{k})a}={\cal D}^{(\underline{0})ab\mu}A^{(\underline{k})b}_\mu-\xi \,m_{(\underline{k})}A_G^{(\underline{k})a},
\end{eqnarray}
where $\xi$ is the gauge--fixing parameter, whose different values correspond to different choices of the gauge.
Using these functions, the gauge--fixing term can be written as
\begin{equation}
{\cal L}_{\rm GF}=\sum_{(\underline{k})}
\left[
\frac{1}{2\xi}A^{(\underline{k})a\mu}{\cal D}^{(\underline{0})ab}_\mu{\cal D}^{(\underline{0})bc}_\nu A^{(\underline{k})c\nu}-m_{(\underline{k})}A^{(\underline{k})a}{\cal D}^{(\underline{k})ab\mu}A_G^{(\underline{k})b}-\frac{1}{2}\xi\,m_{(\underline{k})}^2A_G^{(\underline{k})a}A_G^{(\underline{k})a}
\right].
\label{gftrm}
\end{equation}
The second term of the right--hand side of this equation cancels the only gauge--scalar couplings allowed by the theory, that is, those involving the pseudo--Goldstone bosons $A^{(\underline{k})a}_G$ and which arise from Eq.~(\ref{gsintcurv}).
We will profit from this cancellation later, when we integrate out the Kaluza--Klein excited modes. In general, simplifications are introduced by the elimination of these couplings because it reduces the number of Feynman diagrams in any calculation aimed to derive contributions to some standard Green's function. Since this cancellation was produced by the introduction of covariant gauge--fixing functions, it is clear that the resulting simplifications are a direct consequence of the preservation of gauge symmetry. In other words, the presence of symmetries come along with simplifications in phenomenological calculations.
However, as the third term of Eq.~(\ref{gftrm}) shows, the removal of such unphysical couplings leaves, as a remnant, an unphysical mass term for these spurious scalar degrees of freedom. In the Feynman--'t--Hooft gauge, defined by the condition $\xi=1$, these masses coincide with those of the Kaluza--Klein gauge excited modes.
The first and third terms in Eq.~(\ref{gftrm}) contribute to light Green's functions since the one--loop level and thus are relevant for the present calculation.
\\

The ghost--antighost--fields term, ${\cal L}_{\rm G}$, involves several couplings of Kaluza--Klein ghost fields, $C^{(\underline{k})a}$, and antighost fields, $\bar{C}^{(\underline{k})a}$, with gauge and scalar Kaluza--Klein modes. After inserting the covariant gauge--fixing functions, only two types of these couplings contribute to standard Green's functions at one loop. One of them is a gauge--dependent mass term, whereas the other one is a kinetic term.
\\

We end this section by showing the whole set of couplings that we shall consider in the integration of heavy Kaluza--Klein modes:
\begin{eqnarray}
{\cal L}_{\rm QKK}&=&{\cal L}^{(\underline{0})}_{\rm 4YM}+\frac{1}{2}\sum_{(\underline{k})}A^{(\underline{k})a\mu}\Bigg[\,g_{\mu\nu}{\cal D}^{(\underline{0})ab}_\rho{\cal D}^{(\underline{0})bc\rho}+g_{\mu\nu}\delta^{ac}m_{(\underline{k})}^2-\left( 1-\frac{1}{\xi} \right){\cal D}_\mu^{(\underline{0})ab}{\cal D}_\nu^{(\underline{0})bc}+2gf^{abc}F^{(\underline{0})b}_{\mu\nu} \Bigg]A^{(\underline{k})c\nu}
\nonumber \\ \nonumber \\ &&
-\frac{1}{2}\sum_{(\underline{k})}A^{(\underline{k})a}_{\bar{n}}\left[ {\cal D}^{(\underline{0})ab}_\rho{\cal D}^{(\underline{0})bc\rho}+\delta^{ac}m_{(\underline{k})}^2 \right]A^{(\underline{k})c}_{\bar{n}}
-\frac{1}{2}\sum_{(\underline{k})}A_G^{(\underline{k})a}\left[ {\cal D}^{(\underline{0})ab}_\rho{\cal D}^{(\underline{0})bc\rho}+\delta^{ac}\xi\,m_{(\underline{k})}^2 \right]A_G^{(\underline{k})c}
\nonumber \\ \nonumber \\ &&
+\sum_{(\underline{k})}\bar{C}^{(\underline{k})}\left[ {\cal D}^{(\underline{0})ab}_\rho{\cal D}^{(\underline{0})bc\rho}+\delta^{ac}\xi\,m_{(\underline{k})}^2 \right]C^{(\underline{k})c}+\cdots.
\label{fullolL}
\end{eqnarray}

\section{Gauge--independent integration of Kaluza--Klein excitations}
\label{KKint}
In this section, we carry out the functional integration of all the Kaluza--Klein excited modes and derive the first nonrenormalizable terms~\cite{LLR,BuWy} of an effective Lagrangian expansion governed by the low--energy dynamic variables and symmetries. To this end, we follow the procedure devised by the authors of Ref.~\cite{BiSa}, which was adjusted and implemented, in Ref.~\cite{NTKKint}, to the integration of heavy Kaluza--Klein modes from Yang--Mills theories with one extra dimension.
\\

We begin by defining the effective action, $S_{\rm eff}$, by
\begin{equation}
e^{iS_{\rm eff}}=\int{\cal D}A^{(\underline{k})a\mu}\,{\cal D}A^{(\underline{k})a}_{\bar{n}}\,{\cal D}A_G^{(\underline{k})a}\,{\cal D}\bar{C}^{(\underline{k})a}\,{\cal D}C^{(\underline{k})a}\exp\left\{ i\int d^4x\Big[\, {\cal L}_{\rm 4YM}+{\cal L}_{\rm GF}+{\cal L}_{\rm G} \Big] \right\},
\label{eSeff}
\end{equation}
where $(\underline{k})\ne(0,\ldots,0)$, so that this expression involves solely the functional integration of Kaluza--Klein excited modes. Gaussian integration of all the terms in Eq.~(\ref{fullolL}), according to the definition given in Eq.~(\ref{eSeff}), yields
\begin{eqnarray}
i\,S_{\rm eff}\equiv i\int d^4x\,{\cal L}_{\rm eff}&=&i\,S^{(\underline{0})}_{\rm 4YM}-\frac{1}{2}\sum_{(\underline{k})}{\rm Tr}\,\log i\Bigg[ -g_{\mu\nu}\big({\cal D}^{(\underline{0})}\big)^2+\left( 1-\frac{1}{\xi} \right){\cal D}^{(\underline{0})}_\mu{\cal D}^{(\underline{0})}_\nu+2ig\,F^{(\underline{0})}_{\mu\nu}-m_{(\underline{k})}^2\,g_{\mu\nu}\,\boldsymbol{1}_N \Bigg]
\nonumber \\ \nonumber \\ &&
-\frac{1}{2}\,\sum_{(\underline{k})}\,{\rm Tr}\,\log i \,\boldsymbol{1}_{n-1}\bigg[ \big( {\cal D}^{(\underline{0})} \big)^2\,+m_{(\underline{k})}^2\boldsymbol{1}_N \bigg]-\frac{1}{2}\sum_{(\underline{k})}{\rm Tr}\,\log i\bigg[ \big( {\cal D}^{(\underline{0})} \big)^2+\xi\,m_{(\underline{k})}^2\boldsymbol{1}_N \bigg]
\nonumber \\ \nonumber \\ &&
+\sum_{(\underline{k})}{\rm Tr}\,\log i\bigg[ -\big( {\cal D}^{(\underline{0})}\big)^2-\xi\,m_{(\underline{k})}^2\,\boldsymbol{1}_N \bigg].
\label{SeffTr}
\end{eqnarray}
Before following through, some explanation on this equation is opportune. The symbol Tr stands for a trace over all degrees of freedom, including spacetime points.
Each trace Tr acts also on matrices that live in different spaces and coexist within the arguments of the logarithms in the different terms, where they multiply each other.
For example, look at the second term of the right--hand side of Eq.~(\ref{SeffTr}).
One way of making sense of the argument of its logarithm is by imagining its terms as $(N^2-1)\times(N^2-1)$ block matrices, which come from the gauge group and whose entries are blocks of size $4\times4$, because of the spacetime indices. This means that, for instance, ${\cal D}^{(0)}$ is the SU($N,{\cal M}^2$) covariant derivative in the adjoint representation of the gauge group and in matrix form, and $\boldsymbol{1}_N$ is the $(N^2-1)\times(N^2-1)$ identity matrix in this gauge--group space, whereas $g_{\mu\nu}$ is the $4\times4$ matrix corresponding to the inverse of the metric tensor.
This term of Eq.~(\ref{SeffTr}) contains all the contributions from the Kaluza--Klein gauge excited modes $A^{(\underline{k})a}_\mu$ and, as it can be appreciated, is gauge dependent, since it carries the gauge--fixing parameter $\xi$.
The third and fourth terms of this equation are scalar contributions from physical scalars $A^{(\underline{k})a}_{\bar{n}}$ and pseudo--Goldstone bosons $A^{(\underline{k})a}_G$, respectively, and the fifth term comprehends all contributions from ghost and antighost fields, $C^{(\underline{k})a}$ and $\bar{C}^{(\underline{k})a}$. Note that the global negative sign within the logarithm of the ghost--sector contribution can be relegated to the zero--point energy. The resulting expression is proportional to the fourth term, which comes from the Kaluza--Klein pseudo--Goldstone modes $A^{(\underline{k})a}_G$. Clearly, the ghost--antighost contributions are minus twice times those produced by the pseudo--Goldstone bosons. This relation, which had already been noticed in the 331 model~\cite{MTTR} and in the five--dimensional Yang--Mills theory~\cite{NTKKint}, is a direct consequence of our set of gauge--fixing functions and illustrates the simplifications supported by the presence of gauge symmetry. This result is general, within this gauge--fixing prescription, and remains the same in any one--loop calculation. Of course, both of these unphysical contributions are gauge dependent. Finally, we point out that the argument of the logarithm of the third term includes the object $\boldsymbol{1}_{n-1}$, which is an identity matrix of size $(n-1)\times (n-1)$ that appears because of the presence of $n-1$ physical scalars $A^{(\underline{k})a}_{\bar{n}}$ per each combination $(\underline{k})$ and per each value of the gauge index $a$.

\subsection{The gauge trace}

In this subsection, we derive a result that is not exclusive of Kaluza--Klein theories, but also applies in other gauge formulations. For that reason, only for now we change our notation to avoid any reference to the particular case of extra--dimensional models. Inspired by the method of Ref.~\cite{BiSa}, and following the Appendix of Ref.~\cite{NTKKint}, we consider a generic trace of the form
\begin{equation}
i\,{\rm Tr}\log\Big[ g_{\mu\nu}(-{\cal D}^2-m^2)-U_{\mu\nu} \Big],
\label{gtrinitial}
\end{equation}
where ${\cal D}$ is the covariant derivative, which we assume to be in some representation of a gauge group SU$(N)$, $m$ is some high--energy scale, and $U_{\mu\nu}=U^a_{\mu\nu}T^a$ is an arbitrary block matrix function of spacetime coordinates whose entries are $4\times4$ matrices. Here, the $T^a$ matrices are the generators of the gauge group in whatever representation we choose. As suggested by the authors of Ref.~\cite{BiSa}, we use the notation $A_\mu\equiv-igT^aA^a_\mu$, so that the covariant derivative reads $D_\mu f(x)=\partial_\mu f(x)+[A_\mu,f(x)]$ and the definition of the Yang--Mills curvature, $F_{\mu\nu}=-igT^aF^a_{\mu\nu}$, has the simple form $[{\cal D}_\mu,{\cal D}_\nu]=F_{\mu\nu}$.
\\

Our objective is to solve the trace given in Eq.~(\ref{gtrinitial}) and so derive an effective expansion, ${\cal L}$, of the form
\begin{equation}
{\cal L}=\sum_{k=1}^\infty\sum_{b=1}^{n_k}\frac{\alpha_{k,b}}{m^{2k-4}}{\cal O}_{k,b}=\sum_{k=1}^\infty\frac{c_k}{m^{2k-4}}\sum_{b=1}^{n_k}\gamma_{k,b}\,{\cal O}_{k,b},
\label{Leffgoal}
\end{equation}
where the $\alpha_{k,b}$ dimensionless coefficients have been written as $\alpha_{k,b}=c_k\,\gamma_{k,b}$, with
\begin{equation}
c_k=\left( \frac{m^2}{4\pi\mu^2} \right)^{\frac{D}{2}-2}\frac{D}{(4\pi)^2}\,\Gamma\left( k-\frac{D}{2} \right).
\end{equation}
The procedure involves momentum integrals, which produce ultraviolet divergences. To handle them, we use the dimensional regularization approach~\cite{BiGi}, for which the dimension of spacetime is set as $D$, with $D\to4$, and a mass scale $\mu$, with units of mass, is introduced to correct units in momentum integrals.
The ${\cal O}_{k,b}$, in Eq.~(\ref{Leffgoal}), are field operators whose dimensions are (mass)$^{2k}$, and $n_k$ is the total number of such operators for a fixed $k$.
In the present study we are concerned with effective operators with mass dimensions as large as six, that is, up to $k=3$.
\\

In this method~\cite{BiSa,NTKKint}, two partial answers are obtained by different means and they are then put together in a particular context to determine the general solution. For the first piece, essentially, one has to guess which will be the general structure of the final result. Though the whole set of SU($N$)--invariant operators of mass dimensions 4 and 6 is well known~\cite{LLR,BuWy}, the presence of the general object $U_{\mu\nu}$ complicates this task. The authors of Ref.~\cite{BiSa} provided a set of operators which works well in the absence of matrices associated to the Lorentz group (which we have represented by means of Lorentz indices).
Fortunately, gauge independence sheds light on what to expect from Eq.~(\ref{gtrinitial}), as we shall appreciate when we implement the final result to the Kaluza--Klein gauge theory. According to this criterion, the correct set for $k=1,2,3$ is
\begin{eqnarray}
\sum_{b=1}^{n_1}\gamma_{1,b}\,{\cal O}_{1,b}&=&\gamma_{1,1}\,\frac{1}{D}{\rm tr}_{\rm r}\big\{ U^\mu\hspace{0.0000001cm}_\mu \big\},
\label{setke1}
\\ \nonumber \\
\sum_{b=1}^{n_2}\gamma_{2,b}\,{\cal O}_{2,b}&=&\gamma_{2,1}\,\frac{1}{D^2}{\rm tr}_{\rm r}\big\{ U^\mu\hspace{0.0000001cm}_\mu\,U^\nu\hspace{0.000001cm}_\nu \big\}+\gamma_{2,2}\,{\rm tr}_{\rm r}\big\{ F_{\mu\nu}F^{\mu\nu} \big\},
\label{setke2}
\\ \nonumber \\
\sum_{b=1}^{n_3}\gamma_{3,b}\,{\cal O}_{3,b}&=&\gamma_{3,1}\,\frac{1}{D^3}{\rm tr}_{\rm r}\big\{ U^\mu\hspace{0.0000001cm}_\mu\,U^\nu\hspace{0.0000001cm}_\nu\,U^\rho\hspace{0.0000001cm}_\rho \big\}+\gamma_{3,2}\,\frac{1}{D^2}{\rm tr}_{\rm r}\big\{ {\cal D}_\mu U^\nu\hspace{0.000001cm}_\nu\,{\cal D}^\mu U^\rho\hspace{0.0000001cm}_\rho \big\}+\gamma_{3,3}\,\frac{1}{D}{\rm tr}_{\rm r}\big\{ F_{\mu\nu}U^\rho\hspace{0.00000001cm}_\rho F^{\mu\nu} \big\}
\nonumber \\ \nonumber \\ &&
+\gamma_{3,4}\,{\rm tr}_{\rm r}\big\{ D_\mu F^{\mu\nu}\,D^\rho F_{\rho\nu} \big\}+\gamma_{3,5}\,{\rm tr}_{\rm r}\big\{ F_{\mu\nu}\,F^{\nu\rho}\,F_\rho\hspace{0.000001cm}^\mu \big\},
\label{setke3}
\end{eqnarray}
with the symbol tr$_{\rm r}$ denoting a trace acting on matrices from the SU($N$) representation in which we are working.
\\

For the second piece, we use the expressions of Ref.~\cite{NT5DYM}, according to which the trace in Eq.~(\ref{gtrinitial}) can be written, in terms of a Lagrangian ${\cal L}$, as
\begin{equation}
i\,{\rm Tr}\,\log\Big[ g_{\mu\nu}(-{\cal D}^2-m^2)-U_{\mu\nu} \Big]=\int d^4x\,{\cal L},
\end{equation}
with
\begin{equation}
{\cal L}=\sum_{k=1}^\infty\frac{(-1)^{k+1}}{k}\,{\rm tr}\left\{ i\,\int\frac{d^4p}{(2\pi)^4}\frac{\left[\delta^\alpha\hspace{0.000001cm}_\beta(\Pi^2+2\,\Pi\cdot p)-U^\alpha\hspace{0.000001cm}_\beta\right]^k}{(p^2-m^2)^k} \right\}\boldsymbol{1}.
\label{gengL}
\end{equation}
To get this expression for the Lagrangian ${\cal L}$, the trace over those degrees of freedom that correspond to points of spacetime has been taken, so that in ${\cal L}$ the trace Tr does not appear. Instead, it involves the trace tr, which acts on all the discrete matrices within the momentum integral. These are block matrices of size determined by the gauge group representation, with entries of size $4\times4$. The symbol {\bf 1}, on the other hand, indicates that operators act on the identity. We have also used the definition~\cite{BiSa} $\Pi_\mu=i\,{\cal D}_\mu$.
\\

After solving the integrals for different values of $k$ in the series given in Eq.~(\ref{gengL}), one should get, with the desired accuracy, the effective Lagrangian expansion shown in Eq.~(\ref{Leffgoal}), with all the coefficients $\gamma_{k,b}$ established exactly. The bunch of technical difficulties that may emanate in a calculation of such nature~\cite{DGMP} was attenuated by the authors of Ref.~\cite{BiSa}, who cleverly conceived a shortcut that renders the calculation of the first nonrenormalizable terms a relatively easy task. While they compared their results and found agreement with other works~\cite{Ball}, we will show below that this method works well enough to yield gauge independent results. As we mentioned already, we will follow dimensional regularization~\cite{BiGi} to deal with ultraviolet divergences. To this end, we carry out the replacement $d^4p/(2\pi)^4\to\mu^{4-D}d^Dp/(2\pi)^D$ in momentum integrals.
\\

For the next step, we connect our ingredients: 1) our conjectured field--operator combinations, Eqs.~(\ref{setke1}--\ref{setke3}); and 2) Eq.~(\ref{gengL}), which is a partial calculation of the effective Lagrangian expansion ${\cal L}$.
The idea is that both of these expressions are valid for any field configuration. In particular, the coefficients $\gamma_{k,b}$ are constant numbers~\cite{Wud}, which remain the same in any field configuration and for any choice of the matrix $U_{\mu\nu}$.
We take a particular configuration in which we denote the gauge fields by $A_\mu\big|_{\rm conf.}=A'_\mu$, and which is defined by~\cite{NTKKint,BiSa}
\begin{equation}
\partial_\mu A'_\nu=0, \,\,\,\,\,\, U^\alpha\hspace{0.000001cm}_\beta=-\delta^\alpha\hspace{0.000001cm}_\beta \,A'^2.
\end{equation}
Within this context,
\begin{eqnarray}
\left[ \delta^\alpha\hspace{0.000001cm}_\beta(\Pi^2+\Pi\cdot p)-U^\alpha\hspace{0.000001cm}_\beta \right]^k=(2i\,p\cdot A')^k\,\delta^\alpha\hspace{0.00001cm}_\beta,
\label{scdef}
\end{eqnarray}
so that using the general solution~\cite{BiSa}
\begin{equation}
i\mu^{4-D}\int\frac{d^Dp}{(2\pi)^D}\frac{p^{\mu_1}p^{\mu_2}\cdots p^{\mu_{2k}}}{(\,p^2-m^2)^{2k}}=\frac{(-1)^{k+1}}{(4\pi)^2}\left( \frac{m^2}{4\pi \mu^2} \right)^{\frac{D}{2}-2}\frac{1}{m^{2k-4}}\frac{\Gamma\left( k-\frac{D}{2} \right)}{2^k\,\Gamma(2k)}S^{\mu_1 \mu_2 \ldots \mu_{2k}}_k,
\end{equation}
Eq.~(\ref{gengL}) can be expressed as Eq.~(\ref{Leffgoal}) evaluated in the special configuration, that is, ${\cal L}|_{\rm conf.}$, with
\begin{equation}
\sum_{b=1}^{n_k}\gamma_{k,b}\,{\cal O}_{k,b}\big|_{\rm conf.}=\sum_{b=1}^{n_k}\gamma_{k,b}\,{\cal O}'_{k,b}=\frac{2^k}{(2k)!}\,{\rm tr}_{\rm r}\{ S_k \}.
\label{Oconf}
\end{equation}
The object $S_k^{\mu_1 \mu_2\ldots\mu_{2k}}$, defining $S_k$ by $S_k=S_k^{\mu_1\mu_2\ldots\mu_{2k}}A'_{\mu_1}A'_{\mu_2}\cdots A'_{\mu_{2k}}$, is a totally symmetric $2k$--tensor that is formed by the sum of all possible different products of $k$ metric tensors. The trace tr$_{\rm r}$, in the last expression, only affects matrices corresponding to the representation of SU($N$). Using Eq.~(\ref{Oconf}), one finds that
\begin{eqnarray}
\sum_{b=1}^{n_1}\gamma_{1,b}\,{\cal O}'_{1,b}&=&{\rm tr}_{\rm r}\big\{ A'\hspace{0.000001cm}^2 \big\},
\label{scdcLt1}
\\ \nonumber \\
\sum_{b=1}^{n_2}\gamma_{2,b}\,{\cal O}'_{2,b}&=&\frac{1}{3}\,{\rm tr}_{\rm r}\big\{ (A'\hspace{0.00000001cm}^2)^2 \big\}+\frac{1}{6}\,{\rm tr}_{\rm r}\big\{ (A'_\alpha A'_\beta)^2 \big\},
\label{scdcLt2}
\\ \nonumber \\
\sum_{b=1}^{n_3}\gamma_{3,b}\,{\cal O}'_{3,b}&=&\frac{1}{45}\,{\rm tr}_{\rm r}\big\{ (A'\hspace{0.00000001cm}^2)^3 \big\}+\frac{1}{15}\,{\rm tr}_{\rm r}\big\{ A'\hspace{0.000001cm}^2(A'_\alpha A'_\beta)^2 \big\}+\frac{1}{30}\,{\rm tr}_{\rm r}\big\{ (A'\hspace{0.000001cm}^2A'_\alpha)^2 \big\}
\nonumber \\ \nonumber \\ &&
+\frac{1}{30}\,{\rm tr}\big\{ (A'_\alpha A'_\beta A'^\alpha)^2 \big\}+\frac{1}{90}\,{\rm tr}_{\rm r}\big\{ (A'_\alpha A'_\beta A'_\gamma)^2 \big\}.
\label{scdcLt3}
\end{eqnarray}
Now we specialize Eqs.~(\ref{setke1}--\ref{setke3}) to this particular configuration, then equate the resulting expressions to Eqs.~(\ref{scdcLt1}--\ref{scdcLt3}), and find the exact values of the constants $\gamma_{k,b}$. We calculate the resulting effective Lagrangian expansion to be
\begin{eqnarray}
{\cal L}&=&\frac{1}{(4\pi)^2}\left[\, 1+\Delta_\epsilon-\log\left( \frac{m^2}{\mu^2} \right) \right]m^2\,{\rm tr}_{\rm r}\big\{ U^\mu\hspace{0.000001cm}_\mu \big\}+\frac{1}{(4\pi)^2}\left[\, \frac{1}{2}+\Delta_\epsilon-\log\left( \frac{m^2}{\mu^2} \right) \right]\frac{1}{8}\,{\rm tr}_{\rm r}\big\{ U^\mu\hspace{0.00001cm}_\mu \,U^\nu\hspace{0.0000001cm}_\nu \big\}
\nonumber \\ \nonumber \\ &&
+\frac{1}{(4\pi)^2}\left[ -\frac{1}{2}+\Delta_\epsilon-\log\left( \frac{m^2}{\mu^2} \right) \right]\frac{1}{3}\,{\rm tr}_{\rm r}\big\{ F_{\mu\nu}F^{\mu\nu} \big\}+\frac{1}{(4\pi)^4}\frac{1}{m^2}\Bigg[ -\frac{1}{96}\,{\rm tr}_{\rm r}\big\{ U^\mu\hspace{0.000001cm}_\mu\,U^\nu\hspace{0.000001cm}_\nu\,U^\rho\hspace{0.000001cm}_\rho\big\}
\nonumber \\ \nonumber \\ &&
+\frac{1}{48}\,{\rm tr}_{\rm r}\big\{ {\cal D}_\mu U^\nu\hspace{0.000001cm}_\nu\,{\cal D}^\mu U^\rho\hspace{0.000001cm}_\rho \big\}-\frac{1}{12}\,{\rm tr}_{\rm r}\big\{ F_{\mu\nu}U^\rho\hspace{0.0000001cm}_\rho F^{\mu\nu}\big\}+\frac{1}{15}\,{\rm tr}_{\rm r}\big\{ {\cal D}_\mu F^{\mu\nu}\,{\cal D}^\rho F_{\rho\nu} \big\}-\frac{2}{45}\,{\rm tr}_{\rm r}\big\{ F_{\mu\nu}F^{\nu\rho}F_\rho\hspace{0.000001cm}^\mu \big\}
\Bigg].
\label{gLeffexpl}
\end{eqnarray}
In this expression, ultraviolet divergences are enclosed by
\begin{equation}
\Delta_\epsilon=\frac{2}{\epsilon}-\gamma_{\rm E}+\log4\pi.
\end{equation}

\subsection{Gauge independence of nonrenormalizable terms}
To implement the effective Lagrangian expansion ${\cal L}$, given in Eq.~(\ref{gLeffexpl}), to the contributions of Kaluza--Klein gauge excited modes, completely contained in the second term of the right hand--side of Eq.~(\ref{SeffTr}), we carry out the replacement~\cite{NTKKint} ${\cal D}^{(\underline{0})}_\mu{\cal D}^{(\underline{0})}_\nu\to(1/D)\big({\cal D}^{(\underline{0})}\big)\hspace{0.0000001cm}^2g_{\mu\nu}-(ig/2)F^{(\underline{0})}_{\mu\nu}$, so that this trace is expressed as
\begin{equation}
-\frac{1}{2}\sum_{(\underline{k})}{\rm Tr}\,\log\bigg\{ g_{\mu\nu}\bigg[ -\big({\cal D}^{(\underline{0})}\big)^2-\left( 1-\frac{\alpha}{D} \right)^{-1}m_{(\underline{k})}^2\,\boldsymbol{1}_N \bigg]-\left( \frac{\alpha}{2}-2 \right)\left( 1-\frac{\alpha}{D} \right)^{-1}ig\,F^{(\underline{0})}_{\mu\nu} \bigg\},
\label{gtrrearr}
\end{equation}
with $\alpha=1-1/\xi$.
Written in this form, this gauge trace has an effective Lagrangian expansion that fits Eq.~(\ref{gLeffexpl}). It is very important to keep in mind that the replacement $F_{\mu\nu}\to-igF_{\mu\nu}^aT^a=-igF_{\mu\nu}$ has to be done in ${\cal L}$ before puting this result into effect.
The first, second and third terms of Eq.~(\ref{gLeffexpl}) involve divergencies, within the $\Delta_\epsilon$ factor, and logarithms that in MS--like renormalization schemes, such as $\overline{\rm MS}$, become nondecoupling. Due to antisymmetry of the Yang--Mills curvature $F^{(\underline{0})a}_{\mu\nu}$, with respect to Lorentz indices, the first and second terms of Eq.~(\ref{gLeffexpl}) vanish exactly when this result is implemented to the gauge trace given in Eq.~(\ref{gtrrearr}).
The third term, on the other hand, survives, but its divergent and nondecoupling effects are unobservable.
As it is discussed in Ref.~\cite{GLMNNT}, despite the presence of multiple and infinite Kaluza--Klein sums, this term is absorbed by means of the standard renormalization procedure. Omitting divergent and nondecoupling terms, and assuming that nonrenormalizable terms of mass--dimension 6 are much larger than terms of higher orders\footnote{The ATLAS Collaboration has excluded values of $1/R$ that lie below 850\,GeV, for the case of one universal extra dimension~\cite{ATLASUED,ATLASUED2}. Moreover, the results of Ref.~\cite{ACD} suggest that lower bounds increase if more extra dimensions are assumed. In this sense, contributions from operators of mass--dimension 6, suppressed by $1/(R_j^{-1})^2$, are expected to be more important than mass--dimension 8 terms, which involve a $1/(R_j^{-1})^4$ suppression.}, we find that the effective Lagrangian emerging from the pure--gauge Kaluza--Klein contributions reads
\begin{equation}
{\cal L}_{\rm gauge}\approx\frac{1}{(4\pi)^2}\frac{1}{15}\sum_{(\underline{k})}\frac{1}{m_{(\underline{k})}^2}\left[ \frac{1}{4}\left( 1-\frac{1}{\xi} \right)-1 \right]\bigg[ \frac{g^2}{2}\,{\rm tr}_{\rm a}\big\{ {\cal D}^{(\underline{0})}_\mu F^{(\underline{0})\mu\nu}\,{\cal D}^{(\underline{0})\rho}F^{(\underline{0})}_{\rho\nu} \big\}+\frac{ig}{3}\,{\rm tr}_{\rm a}\big\{ F^{(\underline{0})}_{\mu\nu}F^{(\underline{0})\nu\rho}F^{(\underline{0})\mu}_\rho \big\} \bigg],
\label{gcontrib}
\end{equation}
where the symbol tr$_{\rm a}$ indicates that the traces are taken on gauge group generators $T^a_{\rm a}$, in the adjoint representation of SU($N,{\cal M}^4$).
\\

With minor customizations, the result of Ref.~\cite{BiSa} can be utilized to solve all the remaining traces in Eq.~(\ref{SeffTr}). We take such expression and find that the total contribution from the unphysical pseudo--Goldstone bosons, ghosts and antighosts yields the expansion
\begin{equation}
{\cal L}_{\rm unphy}\approx\frac{g^2}{(4\pi)^2}\frac{1}{15}\sum_{(\underline{k})}\frac{1}{m_{(\underline{k})}^2}\frac{1}{4\xi}\bigg[ \frac{g^2}{2}\,{\rm tr}_{\rm a}\big\{ {\cal D}^{(\underline{0})}_\mu F^{(\underline{0})\mu\nu}\,{\cal D}^{(\underline{0})\rho}F^{(\underline{0})}_{\rho\nu} \big\}+\frac{ig^3}{3}\,{\rm tr}_{\rm a}\big\{ F^{(\underline{0})}_{\mu\nu}F^{(\underline{0})\nu\rho}F^{(\underline{0})\mu}_\rho \big\} \bigg],
\label{unphycontrib}
\end{equation}
where, again, we have neglected terms of mass--dimensions larger than 6. As we did in the case of the gauge trace, here we have omitted terms that are proportional to the trace of $F_{\mu\nu}F^{\mu\nu}$, for they can be absorbed by renormalization and are thus unobservable~\cite{GLMNNT}.
The only sources of gauge dependence are the contributions from Kaluza--Klein gauge excited modes, pseudo--Goldstone bosons, and ghost and antighost Kaluza--Klein fields. As it can be appreciated from Eqs.~(\ref{gcontrib}) and (\ref{unphycontrib}), the contributions from the gauge excited modes differ from those generated by unphysical scalars and ghost--antighost fields by gauge dependent global factors. When these contributions are all added together these global factors are summed like
\begin{equation}
\frac{1}{4}\left( 1-\frac{1}{\xi} \right)-1+\frac{1}{4\xi}=-\frac{3}{4}.
\end{equation}
This explicitly demonstrates that the effective Lagrangian expansion resulting from integrating out all the Kaluza--Klein excited modes is gauge independent, which is achieved by a fine cancellation between all gauge dependent contributions.
\\

The rest of the Kaluza--Klein contributions come from the whole set of physical scalars. Recall that these contributions do not introduce gauge dependence. Gathering these scalar contributions with those from gauge and unphysical Kaluza--Klein modes we find the following effective Lagrangian:
\begin{equation}
{\cal L}_{\rm eff}\approx{\cal L}^{(\underline{0})}_{\rm 4YM}
-\bigg[ \frac{g^2}{(4\pi)^2}\frac{N(n+2)}{60}\,{\rm tr}_{\rm f}\Big\{ {\cal D}^{(\underline{0})}_\mu F^{(\underline{0})\mu\nu}\,{\cal D}^{(\underline{0})\rho}F^{(\underline{0})}_{\rho\nu} \Big\}
+\frac{ig^3}{(4\pi)^2}\frac{N(n+2)}{90}\,{\rm tr}_{\rm f}\Big\{ F^{(\underline{0})}_{\mu\nu}F^{(\underline{0})\nu\rho}F^{(\underline{0})\mu}_\rho \Big\} \bigg]\sum_{(\underline{k})}\frac{1}{m_{(\rm k)}^2}.
\label{Leffinfsums}
\end{equation}
Both nonrenormalizable terms in this result involve a factor $N$, which distinguishes gauge group.
This method to derive effective Lagrangians by the integration of heavy degrees of freedom
produced SU($N,{\cal M}^4$)--invariant operators with traces over generators in the adjoint representation of the gauge group. This can be appreciated, for instance, in Eq~(\ref{gcontrib}), which is written in terms of the trace tr$_{\rm a}$. Nevertheless, we have written the total contribution to ${\cal L}_{\rm eff}$ in terms of the trace tr$_{\rm f}$, which represents a trace over generators in the fundamental representation. In other words, we have replaced $F_{\mu\nu}=F^a_{\mu\nu}T^a_{\rm a}$ by $F_{\mu\nu}=F^a_{\mu\nu}T^a_{\rm f}$, where $T^a_{\rm f}$ represents the SU($N,{\cal M}^4$) generators in the fundamental representation, and this change has come along with the previously mentioned factor $N$. Our result also depends on the number $n$ of extra dimensions. Explicit dependence comes from the contribution of physical scalars $A^{(\underline{k})}_{\bar{n}}$, which involves a global factor $(n-1)$ that adds to factors from all other contributions to produce the $(n+2)$ that appears in Eq.~(\ref{Leffinfsums}). Dependence on the number of extra dimensions is also implicitly located within the sum over combinations $(\underline{k})$ of Kaluza--Klein indices.

\subsection{Regularization of multiple Kaluza--Klein sums and final result}
In general, ultraviolet divergences are induced by loop calculations, though renormalizable formulations set the conditions to get rid of them.
The ultraviolet comportment of loop contributions from Kaluza--Klein excited modes features, besides usual divergences, multiple infinite Kaluza--Klein sums that often produce divergent results.
{\it Continuous divergences}, from integration of 4--momenta, and {\it discrete divergences}, from Kaluza--Klein sums, share a common origin. Discrete divergences arise from the split of extra--dimensional momentum, whose components along the extra dimensions get quantized by compactification, producing Kaluza--Klein sums over Kaluza--Klein masses. In the very special case of only one extra dimension, these sums are~\cite{NTKKint,LMNTotro,GLMNNT} finite Riemann--zeta functions, but as more extra dimensions are considered, multiple sums appear and convergence of results is lost.
\\

Extra--dimensional coupling constants are dimensionful, so that these physical descriptions are not renormalizable.
Kaluza--Klein theories, which provide us with a four--dimensional viewpoint, have their own coupling constants, which turn out to be dimensionless. Moreover, the direct generalization of four--dimensional renormalizable formulations, such as the Standard Model, to descriptions placed in extra dimensions generates Kaluza--Klein theories in which all couplings are renormalizable in the Dyson's sense.
Consequently, loop amplitudes in which continuous and discrete divergences coexist can be renormalized~\cite{GLMNNT}. However, in many cases, there are contributions in which only discrete divergences are present and they cannot be eliminated by renormalization.
Regulators aimed at ultraviolet divergences produced by Kaluza--Klein modes and which preserve higher--dimensional Lorentz and gauge symmetries were proposed, analyzed and implemented in Refs.~\cite{BaDi1,BaDi2,BaDi3}.
In Ref.~\cite{GLMNNT}, a regularization scheme to deal with multiple Kaluza--Klein sums was recently introduced and explored. This approach, which is based on the Epstein--zeta function~\cite{Eps,PoTi,NaPa,Zuck,Hard,Sieg,Glass,Glassotro}, allows one to isolate divergences produced by these infinite sums. 
\\

The Epstein--zeta function, which is a generalization of the Riemann--Zeta function, is defined as
\begin{equation}
\zeta(s;Q)=\sum_{0\ne x\in \mathds{Z}^n}\frac{1}{(x^{\rm T}Q\,x)^s}, \hspace{0.3cm}\mathfrak{R}(s)>\frac{n}{2},
\end{equation}
where $s$ is some complex number. The sum and the denominator within it, written in terms of the vector $x$, its transpose $x^{\rm T}$, and the $n\times n$ matrix $Q$, together represent a multiple sum over integer numbers.
Except for a simple pole at $s=n/2$, the Epstein--zeta function is analytically continued by means of the functional equation
\begin{equation}
\zeta(s;Q)=(\det Q)^{-\frac{1}{2}}\frac{\pi^{2s-\frac{n}{2}}\Gamma\left( \frac{n}{2}-s \right)}{\Gamma(s)}\,\zeta\left( \frac{n}{2}-s;Q^{-1} \right).
\end{equation}
Following the prescription of Ref.~\cite{GLMNNT}, we assume that all extra dimensions are equally sized, that is, $R_j\equiv R$ for any $j=1,\ldots,n$. In this context, the Kaluza--Klein sums in Eq.~(\ref{Leffinfsums}) can be written as
\begin{equation}
\sum_{(\underline{k})}\frac{1}{m_{(\underline{k})}^2}=\frac{R^2}{2}\sum_{l=1}^n
\left(
\begin{array}{c}
n
\\
l
\end{array}
\right)\,\zeta(1;I_l),
\end{equation}
where $I_l$ is the $l\times l$ identity matrix. The isolated singularities of these sums are given by $l=2$. Using these results, we finally write the effective Lagrangian expansion as
\begin{eqnarray}
{\cal L}_{\rm eff}&\approx&{\cal L}^{(\underline{0})}_{\rm 4YM}-\frac{g^2}{(4\pi)^2}\frac{N(n+2)}{120}R^2\sum_{l=1}^n\frac{n!}{l!(n-l)!}\,\zeta(1;I_l)\,{\rm tr}_{\rm f}\Big\{ {\cal D}^{(\underline{0})}_\mu F^{(\underline{0})\mu\nu}\,{\cal D}^{(\underline{0})\rho}F^{(\underline{0})}_{\rho\nu} \Big\}
\nonumber \\ \nonumber \\ &&
-\frac{ig^3}{(4\pi)^2}\frac{N(n+2)}{180}\,R^2\sum_{l=1}^n\frac{n!}{l!(n-l)!}\,\zeta(1;I_l)\,{\rm tr}_{\rm f}\Big\{ F^{(\underline{0})}_{\mu\nu}F^{(\underline{0})\nu\rho}F^{(\underline{0})\mu}_\rho \Big\}.
\end{eqnarray}
These nonrenormalizable terms are divergent at $l=2$, which means that in the case of only one extra dimension ($n=1$) all results are finite. This property of five--dimensional models has been reported in phenomenological studies~\cite{LMMNTT}.
\\

As it is discussed in Ref.~\cite{GLMNNT}, the divergent behavior of loop amplitudes that include infinite Kaluza--Klein multiple sums is determined by the number of extra dimensions and by the accuracy achieved in a given calculation. In the present paper, we have restricted our result to the very first nonrenormalizable terms of the effective Lagrangian expansion, that is, to those with mass--dimension 6.
A more accurate effective Lagrangian, however, would include terms with larger mass dimensions. For instance, assume that we cut the effective Lagrangian series at $1/m^{2K}_{(\underline{k})}$, for some natural number $K$. The resulting effective Lagrangian would read
\begin{equation}
{\cal L}_{\rm eff}={\cal L}^{(\underline{0})}_{\rm 4YM}+\sum_{(\underline{k})}\Bigg[ \sum_{j=1}^K\sum_{a=1}^{N_j}R^{2j}\frac{\beta_{j,a}}{(\underline{k}\hspace{0.0000001cm}^2)^j}\,{\cal O}_{j,a} \Bigg],
\label{genKKeffLexp}
\end{equation}
where the $\beta_{j,a}$ are dimensionless constant coefficients that depend on the gauge group, on the number of extra dimensions, and on the four--dimensional coupling constant $g$. The ${\cal O}_{j,a}$ factors represent nonrenormalizable operators of mass dimension (mass)$^{2j}$ and $N_j$ is the total number of such operators for a fixed mass dimension. The underlined symbol $\underline{k}^2$ stands for any sum of quadratic Kaluza--Klein indices: $k_1^2$, \ldots, $k_n^2$, $k_1^2+k_2^2$, \ldots, $k_{n-1}^2+k_n^2$, \ldots,  $k_1^2+\cdots+k_n^2$. Unless a cutoff for the Kaluza--Klein sums is established, the sum over combinations $(\underline{k})$ is formally divergent if $n>1$.  We have seen that nonrenormalizable terms of mass--dimension 6, which correspond to $j=1$, generate a divergence in $\zeta(1,I_l)$, at $l=2$ .
If somehow mass--dimension 6 terms were not present, we may have cut the seres at $j=2$ (mass--dimension 8). Such terms would involve Epstein--zeta functions $\zeta(2,I_l)$ instead, with a divergence at $l=4$ as long as $n\geqslant4$, though being finite if $n<4$. In general, any nonrenormalizable term with mass--dimension $k$, alone, generates ultraviolet discrete divergences if $n\geqslant k-4$, but such divergences are absent if $n< k-4$.
Thus, if a more accurate effective Lagrangian is derived, a different divergent behavior is expected for a large enough number of extra dimensions. It is important stressing that this discussion is rather qualitative, for there are individual divergent sums $\sum_{(\underline{k})}1/(\underline{k}\hspace{0.000001cm}^2)^j$, which means that using $\sum_{(\underline{k})}\sum_j=\sum_j\sum_{(\underline{k})}$ in Eq.~(\ref{genKKeffLexp}) may lead to incorrect results. Strictly speaking, in presence of nonrenormalizable terms of mass dimension larger than 6 we cannot separately analyze each sum $\sum_{(\underline{k})}1/(\underline{k}\hspace{0.000001cm}^2)^j$, for each different value of $j$, and then infer the divergent nature of the whole effective Lagrangian expansion.

\section{Summary}
\label{DandS}
Assuming that, besides the ordinary four--dimensional coordinates, spacetime comprises a set of compact dimensions, each one with an orbifold structure, we have worked with a gauge theory governed by the SU($N,{\cal M}^{4+n}$) group, in a framework in which extra dimensions are universal. We have integrated out all the four--dimensional heavy degrees of freedom that are produced by this formulation and which are part of the particle spectrum of the corresponding Kaluza--Klein theory. The main result was the derivation of the first nonrenormalizable terms, with mass--dimension 6, of an effective Lagrangian that abides by low--energy symmetries and whose field content is the set of Kaluza--Klein zero modes. In passing, we have studied diverse aspects of extra--dimensional gauge theories.
In particular, we have given a glimpse on the quantization of these Kaluza--Klein descriptions, including an SU($N,{\cal M}^4$)--covariant set of gauge--fixing functions, which preserves part of the gauge invariance of the Kaluza--Klein theory and thus induces valuable simplifications in calculations. For instance, contributions from ghost--antighost fields are related in a simple way to those from pseudo--Goldstone bosons.
We have also provided the full set of couplings that involve heavy modes, including Kaluza--Klein ghost and antighost fields, and which contribute to standard Green's functions since the one--loop level. We have emphasized that our resulting effective Lagrangian is gauge independent with respect to gauge--fixing of the Kaluza--Klein gauge excited modes.
The imperative joint intervention of the contributions from Kaluza--Klein heavy gauge fields, pseudo--Goldstone bosons, and ghost and antighost fields to attain gauge independence was pointed out.
We implemented a novel regularization approach, based on the Epstein--zeta function, which is intended to isolate ultraviolet divergences that are encrypted within multiple and infinite Kaluza--Klein sums. In applying the Epstein--zeta regularization to our results, divergences were explicitly separated and precisely located. The resulting expression shows that, as expected, divergences of the first nonrenormalizable terms only arise when the number of extra dimensions is 2 or greater. We point out that, for a large enough number of extra dimensions, the divergent behavior of a more accurate effective Lagrangian expansion will be modified by nonrenormalizable terms of larger mass dimensions.
To this respect, recall that accuracy, in the context of effective theories, means to add more nonrenormalizable terms. Standard divergencies, from integration of continuous four--dimensional momentum, also appear in our expansion. Moreover, nondecoupling effects driven by logarithmic terms that are subject to multiple Kaluza--Klein sums are present as well. However, we have stressed that the renormalization procedure used to remove standard divergencies eliminates these effects, rendering them unobservable.

\begin{acknowledgments}
The authors acknowledge financial support from CONACYT (M\'exico). 
HNS and JJT also acknowledge financial support from SNI (M\'exico).
\end{acknowledgments}

\end{document}